\documentstyle[sprocl]{article}

\input{psfig}
\def\be{\begin{equation}}
\def\ee{\end{equation}}
\def\bea{\begin{eqnarray}}
\def\eea{\end{eqnarray}}

\begin{document}

\title{RECENT THEORETICAL PROGRESS IN $t\bar{t}$ THRESHOLD
ANALYSES \footnote{Talk given at the
International Workshop on Linear Colliders,
April 28 - May 5, 1999, Sitges, Spain.}$^,$ 
\footnote{TU-582, November 1999.}
}

\author{ Y. Sumino }

\address{Department of Physics, Tohoku University\\
    Sendai, 980-8578 Japan}


\maketitle\abstracts{
I review the theoretical progress made
in these 4 years concerning the process
$e^+e^- \to t\bar{t}$ in the threshold region.
I summarize the progress towards each of the three
major physics 
goals that should be achieved at $t\bar{t}$ threshold region.
}
  
\section{Introduction}

In this short note I summarize the progress in our theoretical
understanding of the process $e^+e^- \to t\bar{t}$ in the threshold region
that has been made
since the last Linear Collider Workshop held in Morioka, 1995.

The major three goals that should be achieved in the operation of
a future $e^+ e^-$ linear collider in the $t\bar{t}$ threshold region 
will be:
\begin{itemize}
\item
Precision measurement of the top quark mass.
\item
Measurements of various top quark interactions:
search for anomalies in
$t\bar{t}\gamma$, $t\bar{t}Z$, $tbW$, $ttg$ and $ttH$ couplings
and possibly for other non-SM interactions.
\item
To elucidate dynamics involved in formations and decays 
of (remnant of) toponium resonances.
\end{itemize}

Regarding the first goal, there has been immense theoretical
progress during last few years; 
towards the second goal, we have paved the way during this
period, but real
theoretical analyses are yet to be done;
as for the third goal, some understanding of the dynamics has been
added.
Below I will review how much progress has been made towards 
each of these goals.
At the same time, new theoretical problems that
are now recognized will be addressed.

\section{The Top Quark Mass}

The top quark mass is one of the fundamental parameters of the SM.
It is conceivable that, in the future,
to know its precise value will be crucial
to have a high predictive power in precision physics.
The uniqueness of top quark is that 
it is the heaviest particle among all the known elementary 
particles.
Hence the mass term of top quark breaks the $SU(2)_L \times U(1)_Y$
symmetry maximally in the SM Lagrangian.
Since the ultimate goal of a linear collider will be
to uncover the mechanism of the symmetry breaking, when we really start
to learn something about it, 
the top quark mass will be among the first clues to search 
into the mystery;
even if the symmetry-breaking sector is described by
that of the SM, knowing precisely the top quark mass translates to knowing
precisely the top Yukawa coupling.

Experimentally the top quark mass will be determined most accurately from
a scan of the $t\bar{t}$ production cross section in the threshold
region.
Previous simulation studies showed that a top mass
measurement with a statistical error of 200~MeV will be 
possible,\cite{exp} so
the task for theorists was to reduce the theoretical uncertainties 
at least to a similar level.
Theorists used to relate the threshold cross section to
the pole mass of top quark, but the pole mass suffers from a large
theoretical uncertainty originating from a renormalon pole.
One of the most impressive theoretical developments of the past
years is conceptual progress in our understanding
of the top quark mass.
It was found \cite{renormalon} 
that the renormalon poles contained in the pole mass
and in the QCD potential
cancel in the total energy of a static quark-antiquark 
($q\bar{q}$) pair
if the pole mass $m_{\rm pole}$ is expressed in terms of
the $\overline{\rm MS}$ mass:
\begin{eqnarray}
&&
E_{\rm tot}(r) = 2 m_{\rm pole} + V(r) ,
\\
&&
V(r) \simeq - \int 
\frac{d^3{\bf q}}{(2\pi)^3} 
\, e^{i {\bf q} \cdot {\bf r}} \, 
C_F \frac{4\pi\alpha_S(|{\bf q}|)}{|{\bf q}|^2} ,
\\
&&
m_{\rm pole} \simeq m_{\overline {\rm MS}}(\mu ) +
\frac{1}{2} {\hbox to 18pt{
\hbox to -5pt{$\displaystyle \int$} 
\raise-15pt\hbox{$\scriptstyle |{\bf q}|< \mu$} 
}}
\frac{d^3{\bf q}}{(2\pi)^3} \, 
C_F \frac{4\pi\alpha_S(|{\bf q}|)}{|{\bf q}|^2} ,
\end{eqnarray}
where the $\overline{\rm MS}$ coupling $\alpha_S(\mu)$
contains a renormalon pole at $\mu = \Lambda_{\rm QCD}$.
The potential $V(r)$ is essentially the Fourier transform of the
Coulomb gluon propagator exchanged between $q$ and $\bar{q}$;
the difference of $m_{\rm pole}$ and $m_{\overline {\rm MS}}$ is 
essentially the infrared portion of the top quark 
self-energy.
The signs of the renormalon contributions are opposite
between $V(r)$ and $m_{\rm pole}$ because the color charges are
opposite between $q$ and $\bar{q}$ while the self-enregy is
proportional to the square of a same charge.
Their magnitudes differ by a factor of two because
both the $q$ and $\bar{q}$ propagator poles contribute in the
calculation of the potential whereas either of the two contributes
in the calculation of the self-energy.
Expanding the Fourier factor $e^{i {\bf q} \cdot {\bf r}}$ 
in a Taylor series for small ${\bf q}$, the leading
renormalon contributions cancel in $E_{\rm tot}(r)$.

As a result of this cancellation, 
the series expansion of the total energy in $\alpha_S(\mu)$ behaves
better if we use the $\overline{\rm MS}$ mass instead of
the pole mass.
This suggests that the $\overline{\rm MS}$ mass has a more natural
relation to
physical quantities of a static (or nonrelativistic) quark-antiquark
system.
Some important aspects are:
\begin{itemize}
\item
In any case the 
pole mass of a quark is ill-defined beyond perturbation theory.
A physically meaningful quantity is the total energy of a color
singlet system. 
Specifically a $t\bar{t}$ pair produced in $e^+e^-$
annihilation is a color singlet state.
\item
When the size of a color singlet system is much smaller than
$\Lambda_{\rm QCD}^{-1}$, infrared gluons with wavelengths 
$\Lambda_{\rm QCD}^{-1}$ cannot couple
to color sources inside the system ---
such a picture is naturally described by the bare QCD Lagrangian.
Hence, if we use $m_{\overline {\rm MS}}$,
which is more closely related to the bare mass than $m_{\rm pole}$ is,
contributions from the infrared gluons vanish in $E_{\rm tot}(r)$.
\item
To describe physical quantities of other processes,
the $\overline{\rm MS}$ mass is more suited 
for an input parameter than the pole mass,
since the renormalon cancellation seems to be a universal feature which
takes place process independently.

\end{itemize}

\begin{figure}
~~~~~~~~~ ~~~~~~~~~~
\psfig{figure=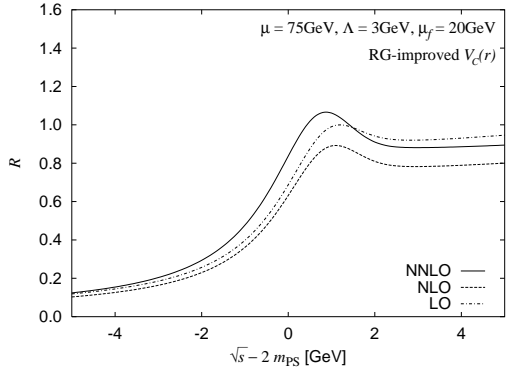,height=5cm}
\caption{The threshold cross section vs.~the energy measured from
the twice the potential-subtracted mass at the scale
20~GeV.  
See the first paper in ref.{\protect \cite{momdist}} for details.
\label{fig1}
}
\end{figure}

Another important theoretical progress is the calculation
of the next-to-next-to-leading order corrections to the threshold
cross section.\cite{NNLO,momdist}
Incorporating the renormalon cancellation, we compare in Fig.~\ref{fig1}
the total cross sections at leading order, next-to-leading order,
and next-to-next-to-leading order.
Combining this study with the relation between 
$m_{\rm pole}$ and $m_{\overline {\rm MS}}$ at ${\cal O}(\alpha_S^3)$
calculated very recently,\cite{cs}
we end up with a theoretical uncertainty of order $200$~MeV
in the relation between the threshold peak position and 
the $\overline{\rm MS}$ mass of top quark 
$m_{\overline {\rm MS}}(m_{\overline {\rm MS}})$;
the uncertainty is dominated by the uncertainty in the present 
value of $\alpha_S(m_Z)$.
With high statistics, the peak position can be measured with
a statistical error of around 50~MeV.\cite{exp2}
Then if we can pin down the value of $\alpha_S(m_Z)$ from other
sources, we may eventually be able to obtain a value of
$m_t$ with an accuracy 50-70~MeV (relative accuracy of $3 \times 10^{-4}$).

It should be noted, however, that there remains fairly large uncertainty
in the normalization of the cross section due to very large
higher order corrections.
I estimate the uncertainty to be 10-15\% at the present stage, and it
remains our task to understand the origin of these large higher order
corrections.
Also, another important task is to calculate the decay widths of 
(would-be) $t\bar{t}$ resonances at next-to-next-to-leading order.

\section{Search for Anomalous Top Interactions}

The fact that the mass term of top quark breaks EW
symmetry maximally suggests that top quark is coupled to the
symmetry-breaking sector more strongly than other particles.
Therefore it is natural to test various top quark interactions
to search for hints to the symmetry-breaking mechanism.
Already several sensitivity studies for anomalous top couplings
have been done in the open top region ($\sqrt{s} \gg 2m_t$).
Certainly it is desirable to test top couplings at highest
possible energy,
where we can explore more deeply into structures of these couplings.
Nevertheless, there are also definite advantages to test top quark
couplings in the $t\bar{t}$ threshold region.
First, to study decay properties of top quark, the threshold region
will be the best place.
This is because:
top quark can be polarized maximally in the threshold region
by polarizing the electron beam (90\% top quark polarization for
80\% longitudinal polarization of $e^-$);
we are guaranteed to be sitting almost at the rest frame of top quark
without need to reconstruct its momentum;
we do not gain resolving power by going to the open top region.
Secondly, we can test not only EW interactions of top quark
but also QCD interactions.
Since the threshold cross section includes an infinite number of
$ttg$ couplings at leading order, it has a high sensitivity also to
anomalous $ttg$ couplings.

Some time ago, sensitivity studies were
performed on the top quark width \cite{exp} and on
the Higgs effects \cite{exp,hjk} in the threshold region. 
Extracting these quantities, however, requires a precise 
theoretical prediction of the 
normalization of the threshold cross section.
In view of the present large theoretical uncertainty, it would be difficult
to extract the Higgs effects. 
For the top width, we will still be able to extract its value with better than
15\% accuracy, and we may extract additional informations from 
measurements of
the forward-backward asymmetry and other observables which are less 
sensitive to the normalization
uncertainty; theoretical studies of these observables are not yet completed,
however.
Studies of sensitivities to more general top couplings have not been
performed yet.
During the past years, next-to-leading order calculations of 
SM contributions to a production-decay chain of top quark
have been carried out.\cite{fsi}
These calculations were needed for any realistic
studies of top couplings, since we need to disentangle different
spin states of top quark using its decay differential distributions.
Moreover, it was shown that the decay process of a free polarized
top quark can be cleanly separated from the boundstate effects.
Now that we are equipped, I hope to see extensive studies of top couplings
by the time of next Linear Collider Workshop.

\section{Dynamics of the $t\bar{t}$ Production and Decay}

A $t\bar{t}$ pair produced below threshold forms
an ideal ``QCD hydrogen-like atom'', and its dynamics is
dictated by perturbative QCD and EW theories in a theoretically
calculable way.
Already, a number of theoretical studies elucidated 
interesting dynamism involved in its time evolution;
especially interplays of QCD and EW interactions make
its properties very unique.
Still these predictions have to be tested; 
for example, nobody has ever compared
a quarkonium wave function calculated from perturbative QCD
with one extracted from experiments, and
this will be possible for the $t\bar{t}$ pair.
Here, I mention new theoretical understanding of 
dynamics that has been added recently.

Firstly, the full next-to-leading order calculations of a
top production-decay chain \cite{fsi}
clarified how the $t$ decay process is affected by
the opposite color charge ($\bar{t}$ or $\bar{b}$) 
sitting nearby.
This effect is particularly enhanced in the threshold region since
$t$ and $\bar{t}$ are produced close to each other
and almost at rest.
Essentially all effects can be understood as originating from
attraction of both parent $t$ and
decay daughter $b$ in the direction of the opposite
charge, but the resulting effects are quite non-trivial since the
$V \! - \! A$ structure of the top decay vertex brings in a correlation between
the top spin direction and directions of emitted decay particles.

Secondly, the next-to-next-to-leading order calculations of
cross sections \cite{NNLO,momdist} incorporated new interactions between
$t$ and $\bar{t}$: the relativistic corrections, the Darwin
potential, the $1/r^2$ potential, etc.
Their net effect enhanced the total cross section 
unexpectedly largely.
On the other hand, their net effect on the shape of top momentum 
distribution \cite{momdist} was shown to be of moderate size
(a few \%).\footnote{
It was shown that the present next-to-next-to-leading order
calculations of top momentum 
distribution are
gauge dependent.\cite{gauge}
}
Since the effects of the $1/r^2$ potential are dominant,
it is understandable qualitatively that 
the total cross section, which is more sensitive to a short
distance region of the potential, is more affected.
Nevertheless, I would very much like to know the physics origin of
the poor convergence of the normalization of
the cross section as we include higher order corrections.

\section*{Acknowledgments}
This work is based on collaborations with T.~Nagano, A.~Ota and 
M.~Peter.
This work was supported by the Japan-German Cooperative Science
Promotion Program.

\end{document}